\documentclass[aps,prb,floatfix]{revtex4}
\usepackage{graphicx}
\newcommand{\BEQ}{\begin{eqnarray}}
\newcommand{\EEQ}{\end{eqnarray}}
\newcommand{\BEA}{\begin{eqnarray}}
\newcommand{\EEA}{\end{eqnarray}}

\renewcommand{\d}{{\rm d}}

\begin{document}

\title{Where Bell went wrong}
% My Copenhagen Interpretation of Bell Inequality Violation}
%\date{Version Aug 29, 2008; today=\today}
%\classification{02.50.Ey, 03.50.DE, 03.65.Ta, 03.65.Ud}


\author{Th. M. Nieuwenhuizen}
\affiliation{Institute for Theoretical Physics,
Valckenierstraat 65, 1018 XE Amsterdam, The Netherlands}

\begin{abstract}
It is explained  on a physical basis how contextuality
allows Bell inequalities to be violated, without bringing an
implication on locality or realism. Hereto we connect first to the
local realistic theory Stochastic Electrodynamics, and then put
the argument more broadly. Thus even if Bell Inequality Violation
is demonstrated beyond reasonable doubt, it will have no say on
local realism.
\end{abstract}

\maketitle

Quantum theory describes in my view the statistics of outcomes of
measurements done on an underlying reality known to us as
``Nature'', while quantum mechanics or quantum field theory should
be called ``a theory'', or ``our present theory''. In my view, in
Nature particles are definite entities, subject to certain waves,
that partly manifest themselves as the mysterious ``quantum
fluctuations''. This view arose from studying the dynamics of
quantum measurements \cite{ABNmeasEPL}, and the subsequent
question of what is going on in an individual quantum measurement.
Some specific process must be going on in every individual
measurement. We have no theory for that, but clearly Nature is
using it, everyday the whole day. Quantum theory gives some
admittedly strong, but incomplete information about outcomes of
experiments.

One may also wonder what is going on in reality with cosmic rays,
particles that have travelled to us for millions of years, or,
even more stunning, with cosmic microwave background radiation,
that travelled more than 13 billion years through empty space.
These are not questions that should be answered within quantum
mechanics, with answers like ``In Hilbert space the state of a
particle is represented by a state vector$\cdots$'', no, they are
questions about what occurs in Nature. I see no other possibility
than to assign a reality to cosmic ray particles and to photons,
say ``balls'' or, preferably, ``solitons'', that travelled all
these years to us through space.

``Quantum fluctuations'' present a notion taught in any quantum
mechanics course without anybody ever explaining {\it what} is
fluctuating {\it where}. Quantum Mechanics and quantum field
theory have the amazing property that we do not have to know these
details if our aim is restricted to getting statistical
predictions. This situation is somewhat reminiscent to the fact
that, given a certain country, we don't have to speak its language
to understand the statistics of its population such as the average
age, average height, average income, and so on, of its people. But
one cannot claim to understand the people without knowing their
language. In other words: statistical understanding (quantum
theory) is partial understanding that should never be taken for
the full truth.

Stochastic Electromagnetism (SED)~\cite{delaPenaCettoBook} is to
this date the most promising option to deal with the underlying
level of reality. In that theory, ``quantum fluctuations'' are
physical fluctuations of the classical electromagnetic field with
a zero-point spectrum. Planck's constant enters by the strength of
these fluctuations. A connection with quantum mechanics has been
put forward by Cetto and de la Pe\~na already some 15 years ago in
their approach called ``Linear Stochastic Electromagnetics'', see
e.g. ~\cite{delaPenaCettoBook,delaPenaCetto}.

With quantum fluctuations expressed by SED or a comparable theory,
and particles being solitons, the underlying quasi-deterministic
level may be called ``Stochastic Soliton Mechanics'', a name I
coined earlier in this series.~\cite{Nh} Double slit interference
should then emerge from solitons going through one of the slits
and interfering with ``idler waves'' originating from the other
slit.

In his opening address of the 2008 V\"axj\"o conference
Foundations of Probability and Physics - 5,  Andrei Khrennikov
took the position that violations of Bell
inequalities~\cite{BellBook} occur in Nature, but do not rule out
local realism, due to contextuality: the measurements
needed to test Bell inequalities (BI) such as the BCHSH inequality
cannot be performed simultaneously ~\cite{Khrennikov}. Therefore
Kolmogorian probability theory starts and ends with having
different probability spaces, and Bell inequality violation (BIV)
just proves that there cannot be a reduction to one common
probability space.  This finally implies that no conclusion can be
drawn on local realism, since incompatible information can not be
used to draw any conclusion. As explained below, the different
pieces of the CHSH inequality involve fundamentally different
distribution functions of the hidden variables, which cannot be
put together in one over all covering distribution of all hidden
variables of the set of considered experiments. To our knowledge,
the first remarks related to contextuality were made by Cetto,
Brody and de la Pe\~na\cite {CettoNonContext}. The contextuality
position was first pointed at in mathematical rigor by Luigi
Accardi~\cite{Accardi}, and then taken by e.g. Fine\cite{Fine},
Pitowsky~\cite{Pitowsky}, Rastal~\cite{Rastal}, Kupczynski
~\cite{Kupczynski}, Garola and Solombrino ~\cite{GarolaSolombrino}
Khrennikov~\cite{Khrennikov02}, Volovich~\cite{Volovich}, Hess and
Philipp~\cite{HessPhilipp}, Sozzo~\cite{GarolaSozzo} and Zhao, de
Raedt and Michielsen~\cite{deRaedt}. Many of their contributions
were reported in previous V\"axj\"o proceedings. I now also
subscribe to this position.

At the University of Amsterdam I supervise bachelor projects on
Bell inequalities. Students are happy to get insight in the
possible structure of the physics behind the quantum formalism.
The role of contextuality is a standard ingredient. Let us see how it
comes up.

\section{How the contextuality enters}

In the Clauser-Horne-Shimony-Holt (CHSH) setup, one may consider a
source that emits pairs of spin 1/2 particles, one going to
station A ``Alice'' and the other, in opposite direction, to
station B ``Bob''. At each station one out of two possible
measurements is performed, $A_1$ or $A_2$ by Alice and $B_1$ or
$B_2$ by Bob. In case $A_1$ the particle's spin is measured along
the axis in direction ${\bf a}_1$ and in case $A_2$ along the axis
in direction ${\bf a}_2$. Likewise $B_{1,2}$ corresponds to
measurements along axis ${\bf b}_{1,2}$ respectively. The
outcomes, ``up" or ``down" along axis ${\bf a}_i$ for A is denoted
as $S_{A_i}=\pm1$, respectively, and likewise for the measurement
by B along axis ${\bf b}_j$ as $S_{B_j}=\pm 1$. The measurement is
repeated many times. Ideally -- if all particles of all pairs are
measured -- the recordings on the two detectors come as pairs. In
each case, the direction of the axis is known and it is recorded
whether ``up'' or ``down'' was measured. Putting afterwards the
results from both detectors together, one determines the four
correlators $C_{ij}=\langle S_{A_i}S_{B_j}\rangle$ for $i=1,2$ and
$j=1,2$ by averaging the outcomes over the pairs. From these four
objects one makes the combination proposed by Clauser, Horne,
Shimony and Holt (CHSH)~\cite{CHSH},

\BEQ \label{BCHSH} BCHSH\equiv C_{11}+C_{12}-C_{21}+C_{22},\EEQ
where ``B'' stands for Bell. Since only $\pm1$ variables are
involved, it will clearly hold that each $|C_{ij}|\le1$ and
$BCSHS\le 4$. But a stronger bound can be derived. Manipulating
with ingredients inside the averages, one has

\BEQ\label{CHSH1} BCHSH=\langle
(S_{A_1}-S_{A_2})S_{B_1}+(S_{A_1}+S_{A_2})S_{B_2}\rangle.\EEQ
Because the $S$ variables are all $\pm1$, one of the two
combinations will be zero, while the other is $\pm 2$. This
implies a version of the Bell inequalities,

\BEQ \label{CHSH2} BCHSH\le 2.\EEQ There are many papers that
investigate in great rigor the validity of the steps made here,
and they again lead to this result.

In the quantum mechanical description of the measurement, the
state is supposed to be pure and described by the state vector
$|\psi\rangle=(|\uparrow_A\rangle|\uparrow_B\rangle-
|\downarrow_A\rangle|\downarrow_B\rangle)/\sqrt{2}$. The
measurement of the particle's spin along axis ${\bf a}_i$ is
described by the operator ${\bf a}_i\cdot\vec\sigma$, where
$\vec{\sigma}=(\sigma_x,\sigma_y,\sigma_z)$ are the three Pauli
matrices with $\sigma_z={\rm diag}(1,-1)$ and we omit the
prefactor $\frac{1}{2}\hbar$. Carrying out the manipulations, one
finds that the maximum over the possible directions is taken when
all vectors are in a plane, ${\bf a}_{1}$ is perpendicular to
${\bf a}_{2}$ and ${\bf b}_{1}$ perpendicular to ${\bf b}_2$,
while the angle between ${\bf a}_{1}$ and ${\bf b}_{1}$ is
$45^\circ$. The value is then $BCSHS= 2\sqrt{2}.$~\cite{CHSH}. In
particular, a value $2<BCSHS<2\sqrt{2}$ is allowed by quantum
mechanics, but violates the Bell inequality (\ref{CHSH2}).

We have not discussed how exactly the measurement is carried out,
only that the results of all pairs were put together. It is
standard to fix $i$ and $j$, say $i=1$, $j=2$ and then to collect
enough measurement outcomes to allow a good statistical analysis.
This is how one uses, say, a neutron beam, during, say, 30
minutes. Next, in a standard setup one changes either $i$ or $j$,
and repeats the measurement during, say, another 30 minutes. In
this way, the four correlators $C_{ij}$ are determined from
consecutive measurements. This setup is good enough to get their
values and to show that the Bell inequality (\ref{CHSH2}) can be
violated. It was applied in the first test of BIV by Freedman and
 Clauser~\cite{FreedmanClauser}.

Bell, however, proposed to choose the measuring directions at A
and B randomly from their two possibilities, at a moment well
after the particles left the source, but well before they arrive
at A and B. It is this selection procedure that brings in the
issue of locality into the problem, that is to say, the question
whether all speeds are less than the speed of light; if not, then
the situation is called non-local. Now if the particles are
separated from each other more than their travel time multiplied
by the speed of light, and detector directions are randomly
chosen, but happen to be in parallel directions, then it appears
always that one of them gives an ``up'' registration and the other
a ``down'' registration. From the point of angular momentum
conservation this is obvious, but it is not obvious how the
particles ``get this done''. Indeed, if the information about the
-- in this case parallel -- directions of the measurement axes is
know only when the particle distance is larger than $ct$, then, it
seems, this information has to be transmitted between them with a
speed larger than $c$. For this reason, Bell's conclusion is that
BIV may point at non-locality. Alternatively, he noticed, it may
be due to the fact that registered values are not related to
properties the particles had before the measurement, a break of
realism.

Either one or the other is broken, so BIV proves, according to
Bell a breakdown of local realism. If this is true, it puts a
major barrier to simple views on the reality underlying quantum
theory and a rather hopeless starting point for attempts to
improve on quantum theory by formulating a subquantum theory.
Absence of local realism is counterintuitive, not to say awkward,
so before giving it up, we should really have not any other
option.

Be it as it may, the first experiment along the lines devised by
Bell was carried out by Aspect, Dalibard and Roger in 1982, thus
consistent with having ruled out local realism~\cite{Aspect82}.
Their work generated a whole field of research, with many
contributions reported in V\"axj\"o meetings. So far, it is agreed
that BIV occurs in many different systems, e.g. for photons,
neutrons, ions and Kaons.

The contextuality issue arises in this discussion because in
definition (\ref{BCHSH}) we have put together correlators that
could not be measured simultaneously. In particular, we have
identified the averaging in the four terms, even though different
things are averaged over. In the standard setup one measures them
in separate runs. In the Bell-Aspect setup one randomly chooses
the directions of the measurement axes, but by the time the
particles arrive, it is set at some choice and from then on it
looks as if it had been in that state all the time. So also then
contextuality plays a role, and in the same way.

\section{Hidden variable models}

\subsection{Bell's hidden variable description}

Bell considers that the measurements outcomes, the  $S_{A_i}=\pm
1$ are determined by some set of hidden variables. Let us denote
the set pertaining to the measured particles schematically by
$\lambda$; being created as a pair, it is supposed that both
particles both share the same set $\lambda$, that travels with
them.
%Since the hidden variables are likely set at the moment
% where the pair is created,
It seems natural to follow Bell and assume

\BEQ C_{ij}=\int\d\lambda \rho(\lambda)S_{A_i}(\lambda)
S_{B_j}(\lambda). \EEQ In this way the four correlators all
involve the same $\rho(\lambda)$. Therefore Eq. (\ref{CHSH2}) can
again be derived from (\ref{BCHSH}) using (\ref{CHSH1}), with
angular brackets now denoting integrals over $\rho(\lambda)$.
Since measurements hint that values $2<BCHSH<2\sqrt{2}$ are
possible, Bell concludes that local hidden variable models do not
work and that Nature lacks local realism.

This argument seems so clear that most in the physics community
are convinced that Bell is right.

\subsection{About loopholes}

Various loopholes are known. The first is the detection loophole
-- in experiments with photons at most 20\% of them are detected.
Such may lead to biases. It was closed in the experiment of Rohe
et al.~\cite{Rohe}. The second is the locality loophole: in
experiments with ions the particles are not well separated, thus
not excluding the possibility of information transfer at speeds
lower than light. It was closed in the experiment of Weihs et
al.~\cite{Weihs}. For neutron double slit interferometry the spin
information can not even in principal be separated from their path
information~\cite{Rauch}. Another case is the coincidence
loophole: when can we speak about the detection of a pair
~\cite{Larsson}. Recently attention was payed to the fair sampling
loophole. So far, fair sampling is a hidden assumption in the
analysis of data, that cannot be checked. G. Adenier has defended
in his PhD-thesis that BIV proves that the fair sampling
assumption is violated, not local realism~\cite{Adenier}.

Since after 25 years since the Aspect experiment it appears still
to be very hard to close all loopholes in a single experiment it
has been supposed, see e.g. Santos~\cite{Santos} that Nature
resists loophole-free Bell experiments. Still, this all is not our
main theme. Our point will be that Bell went wrong even before the
issue of these loopholes has to be addressed, because of the
{\it contextuality loophole}, that cannot be closed.

\subsection{Improved hidden variables description}

Bell's argument would not convince Niels Bohr, since the detectors
have not been taken into account. Clearly, the detectors consist
of many particles and will also have hidden variables,
$\lambda_{A_i}$ and $\lambda_{B_j}$. In this setup, each of the
four correlators can be written as

\BEQ C_{ij}=\int\d\lambda\int\d\lambda_{A_i}\int\d\lambda_{B_j}
\rho_{ij}(\lambda,\lambda_{A_i},\lambda_{B_j})S_{A_i}(\lambda,\lambda_{A_i})
S_{B_j}(\lambda,\lambda_{B_j}), \EEQ where it is to be noted that
we assume that the measured value at A does not involve any
parameter of B, hidden or not. To come back to the steps of Bell,
one has to assume that the four $\rho_{ij}$ arise from one global
distribution function $\rho_G$, so that, for instance,

\BEQ\label{rhoG} \rho_{12}(\lambda,\lambda_{A_1},\lambda_{B_2})
=\int\d\lambda_{A_2}\int\d\lambda_{B_1}
\rho_G(\lambda,\lambda_{A_1},\lambda_{A_2},\lambda_{B_1},\lambda_{B_2}).
\EEQ and likewise for $\rho_{11}$, $\rho_{21}$, $\rho_{22}$. This
is the new way in which contextuality enters: it is assumed that
there exists an underlying distribution $\rho_G$ of the sets of
hidden variables of all the measurements, even though they cannot
be carried out simultaneously.

With this relation, the manipulations that led from (\ref{BCHSH})
to (\ref{CHSH2}) can be repeated and the same Bell inequality can
be derived. It being violated in experiment, one then concludes
that Bell was basically right, the detectors bring no new
information, and hidden variable models are excluded. The
remaining focus is then to close the loopholes and prove that Bell
was indeed right.

\section{Quantum measurements}

One point overlooked by Bell is thus the role of the hidden
variables of the detectors. Contrary to Bohr's philosophy, Bell
did not include them explicitly in his considerations. This may be
blamed to an attitude created by the projection postulate of
quantum mechanics. In textbooks it is mostly postulated that a
quantum measurement amounts to a non-unitary projection of the
quantum state on an eigenstate of the measured operator. This
idealized notion is completely different from what is common
practice in experimental laboratories. Indeed, quantum
measurements are dynamical processes too that can be described
within quantum mechanics.

Indeed, it has been possible to consider a rich enough model \cite{ABNmeasEPL} for
an apparatus that can perform the measurement of a spin
$\frac{1}{2}$. The apparatus is an Ising magnet, consisting of a
large number of quantum spins $\frac{1}{2}$, coupled to each other
only via their z-components (Ising character) and also coupled to
a bath. The magnet starts out in a metastable paramagnetic state,
and by coupling to the tested spin, the magnetization is driven
into its stable up- or down ferromagnetic state. Here the
metastability offers a multiplication of the weak quantum signal
of the tested spin into the macroscopic, stable up- or down value
of the magnetization at the end of the measurement, that is easily
recorded. The bath is also needed, namely for dumping the excess
(free) energy from the initial state of high (free) energy. In
this model, the Schr\"odinger cat states disappear quickly, first
by an NMR-type dephasing due to the interaction of the tested spin
with the spins of the magnet, and then, in the dephased situation,
all memory of the initial state is erased by decoherence due to
the coupling to the bath \cite{ABNmeasEPL}.

This approach thus describes a quantum measurement as a specific
process of quantum mechanics, in which two important timescales
are concerned: the small dephasing time and the somewhat larger
decoherence time of the off-diagonal terms, and the larger
registration time of the diagonal elements, that is, the time in
which the up- or down magnetization is built up. If all are still
rather small, one may consider these processes effectively as
``instantaneous'' and the collapse as a non-unitary evolution.
This is what is taught in most textbooks, and we stress that to an
extent it is misleading. The collapse view holds only in an
effective sense, in reality the complete dynamics is unitary in
the full Hilbert space of tested system and the apparatus. From
the point of view of the tested system, it is an open system
dynamics.

\subsection{Where did Bell make the error?}

The error of Bell lies in the assumption of the existence of a
$\rho_G$, see Eq. (\ref{rhoG}). This was not explicit, since Bell
did not include the role of detectors in his formulas. It was an
implicit error, due to the hidden assumption that detectors could
be left out in the final argument.

But it is absolutely not true that, if one knows the marginals,
here the $\rho_{ij}(\lambda,\lambda_{A_i},\lambda_{B_j})$ for
$i=1,2$ and $j=1,2$, one may conclude that a common distribution
$\rho_G$ exists. There are theorems on this and there are explicit
examples in which some probabilities then have to be negative
~\cite{Vorobev}. The latter does not make sense, so it is better
to say that a common $\rho_G$ does not exist. Physically this is
not a complete surprise, because anyhow the relevant experiments
could not be carried out simultaneously. This uncomfortable
knowledge thus appears to express itself also by absence of a
common probability distribution (mathematicians say: absence of a
common probability space).

\section{Stochastic Electrodynamics}

So far, so good, the above is common knowledge -- even though not
commonly accepted. On my way back from the V\"axj\"o 2008
conference {\it Foundations of Probability and Physics-5} to my
hometown Amsterdam, I realized at the airport of Copenhagen -- it
had to be there -- that a physical argument can be brought into
the discussion of hidden variable theories such as SED and alikes.
In such theories there are specific hidden variables, those that,
at some initial time,  set the stochastic forces acting on the
measurement apparatuses. A different setting of an apparatus
corresponds to a physically different situation and thus to
physically different sets of these hidden variables. In each
setting, they drive the quantum working of the relevant apparatus,
including opposite outcomes when members of a pair are measured
along parallel axes. There is no physical reason why for different
apparatuses the hidden variables should have the same nature, that
is, have a common distribution, that is, be defined on a common
probability space.

This can be made more explicit by imagining that when Alice's
detector is in direction ${\bf a}_1$, there will be put some other
apparatus in direction ${\bf a}_2$. It is immaterial what this is
exactly doing, but for sure it will be driven by the hidden
variables that would drive Alice's detector were it in this
direction. Now it is clear that we speak about {\it physically
exclusive situations}, each setup $A_{1,2}$ is distinctive and it
excludes the other, $A_{2,1}$: {\it One can't have the cake and
eat it.} Again, for this very reason there is no justification to
assume that their hidden variables are described by a common
distribution $\rho_G$.

In any hidden variable theory, one may expect emission of
radiation by Lorentz damping, i. e. accelerated electrons. This is
a physical effect, which in SED is statistically balanced by the
stochastic forces to reach an equilibrium ``quantum'' state, due
to the presence of a fluctuation-dissipation
theorem\cite{delaPenaCettoBook}. This brings once more a physical
aspect of detectors, once more precluding attempts to put
different setups together. Even if this Lorentz damping is not
measurable nowadays, we only need to think of the heating of air
by the apparatus (and by the second, irrelevant one, if it is
there), surely a measurable effect, to realize that {\it different
setups of the detectors exclude each other and thus have no cause
for possessing a common hidden variables distribution.}

Within SED there is a clear understanding of the Bell-Aspect
switching of detector directions: this just has no influence. What
counts is the position of the detector at the moment when the
particle arrives, not what happens before. Freedman and Clauser
already employed this fact when considering detectors without
random switching~\cite{FreedmanClauser}.

David Mermin has formulated a pedagogic model where the members of
the particle pairs carry instruction sets for the outcomes of the
detectors, which contradicts this conclusion~\cite{Mermin}.
However, Adenier showed that his model can reproduce quantum
results if non-detection events are included in the instruction
sets~\cite{AdenierMermin}.

\section{Conclusion}

Violations of Bell inequalities occur in Nature if loophole-free
situations can be reached. The BIV of quantum physics are
adequately explained by quantum theory. Though loopholes have yet
not been closed, we would be very surprised if quantum theory
would not give the right answer. Indeed, if it did not, then how
could it work so well otherwise?

It has always stunned me that Bell's simple hidden variables
argument could have such profound implications as the absence of
local reality. With some experience in deriving physical results
to explain observations in various fields, the Bell analysis has
always appeared rather abstract (mathematical) and suspiciously
simple to me. The above concrete step of a physical implementation
of the contextuality argument makes clear to me on a physical
basis that Bell just overlooked a mathematical issue. The
contextuality argument puts his conclusion where it should be: a
mathematical derivation devoid of a clear physical mechanism, that
can be refuted on the basis of proper mathematics, contextuality,
or, as we showed, on the basis of physics, exclusiveness of
different detector setups.

So far, in literature it is claimed only that a violation of Bell
inequalities leads to absence of a common distribution. Our
physical argument makes clear that it must also be absent when the
BI are not violated.

Assuming a common distribution function for hidden variables of
incompatible experimental setups is like adding apples to oranges.
It is know that two apples plus three oranges do not add up to
five bananas. Likewise, even when combining the outcomes of
results of incompatible setups does lead to results described by
quantum theory, this managing of data does not yield information
about deep physical properties such as locality or realism. The
physical input is much too poor to address those physical
questions. They are, in my view, out of sight of the progress in
physics that we may hope for in next decades.

Bell inequalities are of profound physical interest, as ever, but
they have no say on local reality. Experimental tests of non-local
realism, though reported in Ref. ~\cite{Zeil} in connection with
Bell inequalities, are actually far beyond the present level of
understanding and manipulation. Nature may possess local realism
or not, Bell inequalities have no say on that. For now we can just
keep our cards on the familiar assumption of Nature possessing
local realism.

As for searching the local reality underlying quantum theory, I
conclude that Bell has kept us nicely busy, by obscuring the goal.
We shall gratefully forgive him, he asked important questions and
his efforts led to new fields such as quantum communication and
quantum cryptography. But abstract mathematical reasoning has a
faint chance to capture relevant physical mechanisms, and once
again this was the lesson to learn. Now it is time to get physics
back to the forum of particles, forces and hidden variables. It is
really time to move on!

\end{document}